\documentclass[12pt]{article}

\usepackage{scicite}
\usepackage{times}

\usepackage{amsmath}
\usepackage{url}
\usepackage{longtable}
\usepackage{amssymb}
\usepackage{graphicx}
\usepackage{deluxetable}
\usepackage{pdflscape}
\usepackage{multirow}
\usepackage{color}

\def\lsim{\hbox{\rlap{\raise 0.425ex\hbox{$<$}}\lower 0.65ex\hbox{$\sim$}}}
\def\gsim{\hbox{\rlap{\raise 0.425ex\hbox{$>$}}\lower 0.65ex\hbox{$\sim$}}}
\def\arcmin{\hbox{$^\prime$}}
\def\arcsec{\hbox{$^{\prime\prime}$}}

\newcommand{\apj}{ApJ}

\newcommand{\nat}{Nature}

\newcommand{\procspie}{Proc.\ SPIE}

\newcommand{\apjBib}{Astrophys.\ J.\ }
\newcommand{\mnrasBib}{Mon.\ Not.\ R.\ Astron.\ Soc.\ }
\newcommand{\araaBib}{Annu.\ Rev.\ Astron.\ Astrophys.\ }
\newcommand{\physrepBib}{Phys.\ Rep.\ }
\newcommand{\prdBib}{Phys.\ Rev.\ D}
\newcommand{\paspBib}{Publ.\ Astron.\ Soc.\ Pac.\ }
\newcommand{\ajBib}{Astron.\ J.\ }
\newcommand{\aapsBib}{Astron.\ Astrophys.\ Suppl.\ Ser.\ }

\def\ucsc{1}
\def\car{2}
\def\ifa{3}
\def\nat{4}
\def\lbl{5}
\def\berk{6}
\def\uc{7}
\def\dark{8}
\def\stsci{9}
\def\jhu{10}

\newenvironment{sciabstract}{%
\begin{quote} \bf}
{\end{quote}}

\title{Swope Supernova Survey 2017a (SSS17a),\\ the Optical Counterpart to a Gravitational Wave Source}

\author
{D.~A.~Coulter$^{\ucsc}$,
R.~J.~Foley$^{\ucsc}$,
C.~D.~Kilpatrick$^{\ucsc}$,\\
M.~R.~Drout$^{\car}$,
A.~L.~Piro$^{\car}$,
B.~J.~Shappee$^{\car,\ifa}$,
M.~R.~Siebert$^{\ucsc}$,
J.~D.~Simon$^{\car}$,\\
N.~Ulloa$^{\nat}$,
D.~Kasen$^{\lbl,\berk}$,
B.~F.~Madore$^{\car,\uc}$,
A.~Murguia-Berthier$^{\ucsc}$,\\
Y.-C.~Pan$^{\ucsc}$,
J.~X.~Prochaska$^{\ucsc}$,
E.~Ramirez-Ruiz$^{\ucsc,\dark}$,
A.~Rest$^{\stsci,\jhu}$,\\
C.~Rojas-Bravo$^{\ucsc}$
\\}

\date{}

\begin{document} 


\baselineskip24pt


\maketitle 

\noindent
\normalsize{$^{\ucsc}$Department of Astronomy and Astrophysics, University of California, Santa Cruz, CA 95064, USA}\\
\normalsize{$^{\car}$The Observatories of the Carnegie Institution for Science, 813 Santa Barbara Street, Pasadena, CA 91101}\\
%\normalsize{$^{\mar}$Hubble and Carnegie-Dunlap Fellow}\\
\normalsize{$^{\ifa}$ Institute for Astronomy, University of Hawai'i, 2680 Woodlawn Drive, Honolulu, HI 96822, USA}\\
%\normalsize{$^{\ben}$ Hubble and Carnegie-Princeton Fellow}\\
\normalsize{$^{\nat}$Departamento de F\'{i}sica y Astronom\'{i}a, Universidad de La Serena, La Serena, Chile}\\
\normalsize{$^{\lbl}$Nuclear Science Division, Lawrence Berkeley National Laboratory, Berkeley, CA 94720, USA}\\
\normalsize{$^{\berk}$Departments of Physics and Astronomy, University of California, Berkeley, CA 94720, USA}\\
\normalsize{$^{uc}$Department of Astronomy and Astrophysics, The University of Chicago, 5640 South Ellis Avenue, Chicago, IL 60637}\\
\normalsize{$^{\dark}$Dark Cosmology Centre, Niels Bohr Institute, University of Copenhagen, Blegdamsvej 17, 2100 Copenhagen, Denmark}\\
\normalsize{$^{\stsci}$Space Telescope Science Institute, 3700 San Martin Drive, Baltimore, MD 21218}\\
\normalsize{$^{\jhu}$Department of Physics and Astronomy, The Johns Hopkins University, 3400 North Charles Street, Baltimore, MD 21218, USA}\\
\\
\normalsize{$^\ast$To whom correspondence should be addressed; E-mail:  dcoulter@ucsc.edu.}


\begin{sciabstract}
On 2017 August 17, the Laser Interferometer Gravitational-wave Observatory (LIGO) and the Virgo interferometer detected gravitational waves emanating from a binary neutron star merger, GW170817.  Nearly simultaneously, the {\it Fermi} and INTEGRAL telescopes detected a gamma-ray transient, GRB~170817A.  10.9 hours after the gravitational wave trigger, we discovered a transient and fading optical source, Swope Supernova Survey 2017a (SSS17a), coincident with GW170817. SSS17a is located in NGC~4993, an S0 galaxy at a distance of 40 megaparsecs. The precise location of GW170817 provides an opportunity to probe the nature of these cataclysmic events by combining electromagnetic and gravitational-wave observations.
\end{sciabstract}

Merging binary compact objects such as black holes (BHs) and neutron stars (NSs) are expected to be gravitational wave (GW) sources in the $10$--$10^{4}$~Hz frequency range \cite{Thorne87} that can be observed using interferometers.  The Laser Interferometer Gravitational-wave Observatory (LIGO) recently used this method to detect several binary BH (BBH) mergers \cite{Abbott16:gw,Abbott16:GW151226,Abbott17}.  These discoveries have unveiled a population of relatively massive black holes, tested General Relativity, and led to insights regarding stellar evolution and binary populations \cite{Abbott16:bbh,Abbott16:gr}.  Although it is unlikely that BBH systems produce a luminous electromagnetic (EM) signature, detecting an EM counterpart to a GW event would greatly improve our understanding of the event by providing a precise location and insight into the merger products.  Unlike BBH mergers, binary NS (BNS) mergers are expected to produce gravitationally unbound radioactive material that is visible at optical and infrared wavelengths (a kilonova) \cite{Li98,Metzger10,Roberts11,Barnes16} and perhaps relativistic jets seen as short gamma-ray bursts (SGRBs) \cite{Lee07,Berger14}.  BNS mergers should produce transient, temporally coincident GWs and light.  This has many advantages in comparison to detecting GWs alone, such as possibly constraining the nuclear equation of state, measuring the production of heavy-elements, studying the expansion of the Universe, and generating a clearer picture of the merger event \cite{Freiburghaus99,Lattimer00,Dalal06}.

On 2017 August 17, LIGO/Virgo detected a strong GW signal consistent with a BNS merger, GW170817 \cite{GCN21509}.  A preliminary analysis of the GW data suggested that the two component masses were small enough to be a BNS system.  This event had a low false-alarm rate of 1 per 10,000 years, a 90-percent chance of being localized to an area of 31~deg$^{2}$ (Figs.~1 \& 2), and a distance of $D = 40 \pm 8$~megaparsecs (Mpc) \cite{GCN21509,GCN21513}.  Contemporaneously, the {\it Fermi} and INTErnational Gamma-Ray Astrophysics Laboratory (INTEGRAL) gamma-ray telescopes detected a SGRB both spatially and temporally coincident with the GW event, GRB170817A.  However, the {\it Fermi}/INTEGRAL localization area was larger than the LIGO/Virgo localization area \cite{GCN21506,GCN21507}.

Our One-Meter, Two-Hemisphere (1M2H) collaboration uses two 1-m telescopes, the Nickel telescope at Lick Observatory in California and the Swope telescope at Las Campanas Observatory in Chile, to search for EM counterparts to GW sources.  Our strategy involves observing previously cataloged galaxies whose properties are consistent with the GW data to search for new sources.  This strategy is particularly effective for nearby events with a small distance uncertainty, which reduces the surface density of viable targets \cite{Gehrels16}.  We observe in either $i^{\prime}$ or $i$ band filters (Nickel and Swope, respectively) because those are the reddest bands available on those telescopes and theoretical models predicted that kilonova light curves would be particularly red \cite{Barnes16}.

As the center of the localization region was in the Southern hemisphere and relatively close to the Sun, the Nickel telescope could not observe the GW170817 localization region.  For GW170817, we were able to also use both Magellan telescopes as part of the search [see \cite{SI} for details], allowing a multi-wavelength campaign covering $giH$ bands.  At the time of the trigger, the local time in Chile was 9:41 am (when the Sun was above the horizon), so observations could not begin for more than 10 hours.  Because of the GW position, the majority of the 90-percentile localization region was expected to only be accessible for the first 2 hours after civil twilight that evening (Fig.~1).

%Materials and methods are available as supplementary materials on Science Online.
Using a catalog of nearby galaxies and the three-dimensional GW localization of GW170817 [e.g., \cite{SI}], we created a prioritized list of galaxies in which the source of the GW event could reside (Table~S1).  Our prioritization algorithm includes information about the stellar mass and star-formation rate of the galaxy.  We examined the positions of the 100 highest-priority galaxies to see if multiple galaxies could fit in a single Swope image (field of view of 29.7~arcminutes $\times$ 29.8~arcminutes), so that we could cover the probable locations as efficiently as possible.  We were able to combine 46 galaxies in a total of 12 images (Fig.~2).  The remaining galaxies on the initial list were sufficiently isolated to require their own images.  We designed an observing schedule that first observed the 12 positions covering multiple galaxies, followed by individual galaxies in order of their priority while they were approximately 19.5 degrees above the horizon (corresponding to an airmass of 3.0).

Starting at 23:13 UT, when nautical twilight ended (Sun $>$ 12 degrees below the horizon), 45 minutes after sunset, and ten hours after the GW trigger, we began observing the GW170817 localization region with an $i$-band filter.  The 60-second exposures had a point-source limiting magnitude of $20.0$~mag, corresponding to an absolute magnitude $M_{i}$ of $-13.0$~mag at a distance of $D = 40$~Mpc (uncorrected for foreground Milky Way extinction).  We immediately transferred, reduced, and examined each image by eye.  In the ninth image (Fig.~3), which was initiated at 23:33 UT and contained two high-priority targeted galaxies, we detected an $i = 17.476 \pm 0.018$~mag source that was not present in archival imaging (Fig.~4). We designate the source as Swope Supernova Survey 2017a (SSS17a); it is located at right ascension $13^{\text{h}}09^{\text{m}}48^{\text{s}}.085\pm0.018$, declination $-23^{\circ}22\arcmin53\arcsec.343\pm0.218$ (J2000 equinox).  SSS17a is offset 10.6\arcsec\ (corresponding to 2.0~kpc at 40~Mpc) from the nucleus of NGC~4993, an S0 galaxy at a redshift of 0.009680 \cite{Jones09} and a Tully-Fisher distance of 40~Mpc \cite{Freedman01}.  NGC~4993 was the twelfth most likely host galaxy based on our algorithm, with a 2.2\% probability of being the host galaxy (see Table S1).

After confirming that SSS17a was not a previously known asteroid or supernova (SN) , we triggered additional follow-up observations \cite{Drout17,Pan17,Shappee17} and disseminated our discovery through a LIGO-Virgo Collaboration (LVC) Gamma-ray Coordination Network (GCN) circular [\cite{GCN21529}, see \cite{SI} for details].  We quickly confirmed SSS17a in a Magellan image performing a similar galaxy-targeted search [\cite{GCN21551}, \cite{SI}].  Several other teams also detected the presence of the new source after our original discovery image [see \cite{MMA} for a complete list].  We observed an additional 45 fields after identifying the new source, acquiring 54 images over 3.5 hours and covering 95.3\% of the total probability (as determined by our algorithm) and 26.9\% of the two-dimensional localization probability.  Comparing to Swope images obtained 18--20~days after the trigger, we found no transient objects other than SSS17a in either set of images.  Most galaxies are about $\sim$7\arcmin\ from the edge of a Swope image (1/4 the size of the field of view), corresponding to $\sim$80~kpc at 40~Mpc.  For these regions covered by our images, we can exclude another luminous transient from being associated with GW170817 at the 95.3\% confidence level [e.g., \cite{SI}].  SSS17a is unlike any transient found by SN searches to date, making it an unusual discovery if unassociated with an extraordinary event such as GW170817.  Additionally, SN rates imply we would expect only 0.01~SNe~year$^{-1}$ in the LVC localization volume.  The combination of all available data further indicates that SSS17a is physically associated with GW170817 (probability of a chance coincidence is $\sim${}$10^{-6}$) \cite{GCN21557,Siebert17}.

Our observations were made with a 1-m telescope with an approximately quarter square degree field-of-view camera.  This is in contrast to the strategy of using wide-field cameras, often on larger-aperture telescopes to observe the entire localization region, unguided by the positions of known galaxies \cite{Smartt16,Soares-Santos16}.  While wide-field imagers might be necessary to discover an EM counterpart in a larger localization error region or in a low-luminosity galaxy, such instrumentation was not necessary for the case of GW170817/SSS17a.  Nearly every optical observatory has an instrument suitable for our strategy; even some amateur astronomers have sufficient instrumentation to perform a similar search.  While aperture and field of view are key capabilities in the EM follow-up of future GW sources at the LIGO/Virgo detection limits, when it comes to finding the closest and scientifically fruitful sources like GW170817/SSS17a, the more import factors are telescope location and observational strategy.

\newpage

\section*{Acknowledgments}
We thank the LIGO/Virgo Collaboration, and all those who have contributed to gravitational wave science for enabling this discovery.
We thank J.\ McIver for alerting us to the LVC circular.
We thank J.\ Mulchaey (Carnegie Observatories director),  L.\ Infante (Las Campanas Observatory director), and the entire Las Campanas staff for their extreme dedication, professionalism, and excitement, all of which were critical in the discovery of the first gravitational wave optical counterpart and its host galaxy as well as the observations used in this study.  We thank I.\ Thompson and the Carnegie Observatory Time Allocation Committee for approving the Swope Supernova Survey and scheduling our program.
We thank the University of Copenhagen, DARK Cosmology Centre, and the Niels Bohr International Academy for hosting D.A.C., R.J.F., A.M.B., E.R., and M.R.S.\ during the discovery of GW170817/SSS17a.  R.J.F., A.M.B., and E.R.\ were participating in the Kavli Summer Program in Astrophysics, ``Astrophysics with gravitational wave detections.''  This program was supported by the the Kavli Foundation, Danish National Research Foundation, the Niels Bohr International Academy, and the DARK Cosmology Centre.

The UCSC group is supported in part by NSF grant AST--1518052, the Gordon \& Betty Moore Foundation, the Heising-Simons Foundation, generous donations from many individuals through a UCSC Giving Day grant, and from fellowships from the Alfred P.\ Sloan Foundation (R.J.F.), the David and Lucile Packard Foundation (R.J.F.\ and E.R.) and the Niels Bohr Professorship from the DNRF (E.R.).
A.M.B.\ acknowledges support from a UCMEXUS-CONACYT Doctoral Fellowship.
M.R.D.\ is a Hubble and Carnegie-Dunlap Fellow.  M.R.D.\ acknowledges support from the Dunlap Institute at the University of Toronto.
B.F.M.\ is an unpaid visiting scientist at the University of Chicago
and an occasional consultant to the NASA/IPAC Extragalactic Database.
J.X.P.\ is an affiliate member of the Institute for Physics and
Mathematics of the Universe.
J.D.S.\  acknowledges support from the Carnegie Institution for Science.
Support for this work was provided by NASA through Hubble Fellowship
grants HST--HF--51348.001 (B.J.S.) and HST--HF--51373.001 (M.R.D.) awarded by the Space Telescope Science Institute, which is operated by the Association of Universities for Research in Astronomy, Inc., for NASA, under contract NAS5--26555.

This paper includes data gathered with the 6.5 meter Magellan Telescopes located at Las Campanas Observatory, Chile.
This research has made use of the NASA/IPAC Extragalactic Database (NED) which is operated by the Jet Propulsion Laboratory, California Institute of Technology, under contract with the National Aeronautics and Space Administration.
Figure 4A is based on observations made with the NASA/ESA Hubble Space Telescope,
obtained from the Data Archive at the Space Telescope Science
Institute (https://archive.stsci.edu; Program 14840), which is operated by the Association of Universities for
Research in Astronomy, Inc., under NASA contract NAS 5--26555. These
observations are associated with programs GO--14840.

The data presented in this work and the code used to perform the
analysis are available at https://ziggy.ucolick.org/sss17a/ .

\clearpage

 \begin{figure}
 \begin{center}
 %% Update graphic with RA/Dec axis labels
 \includegraphics[angle=0,width=6in]{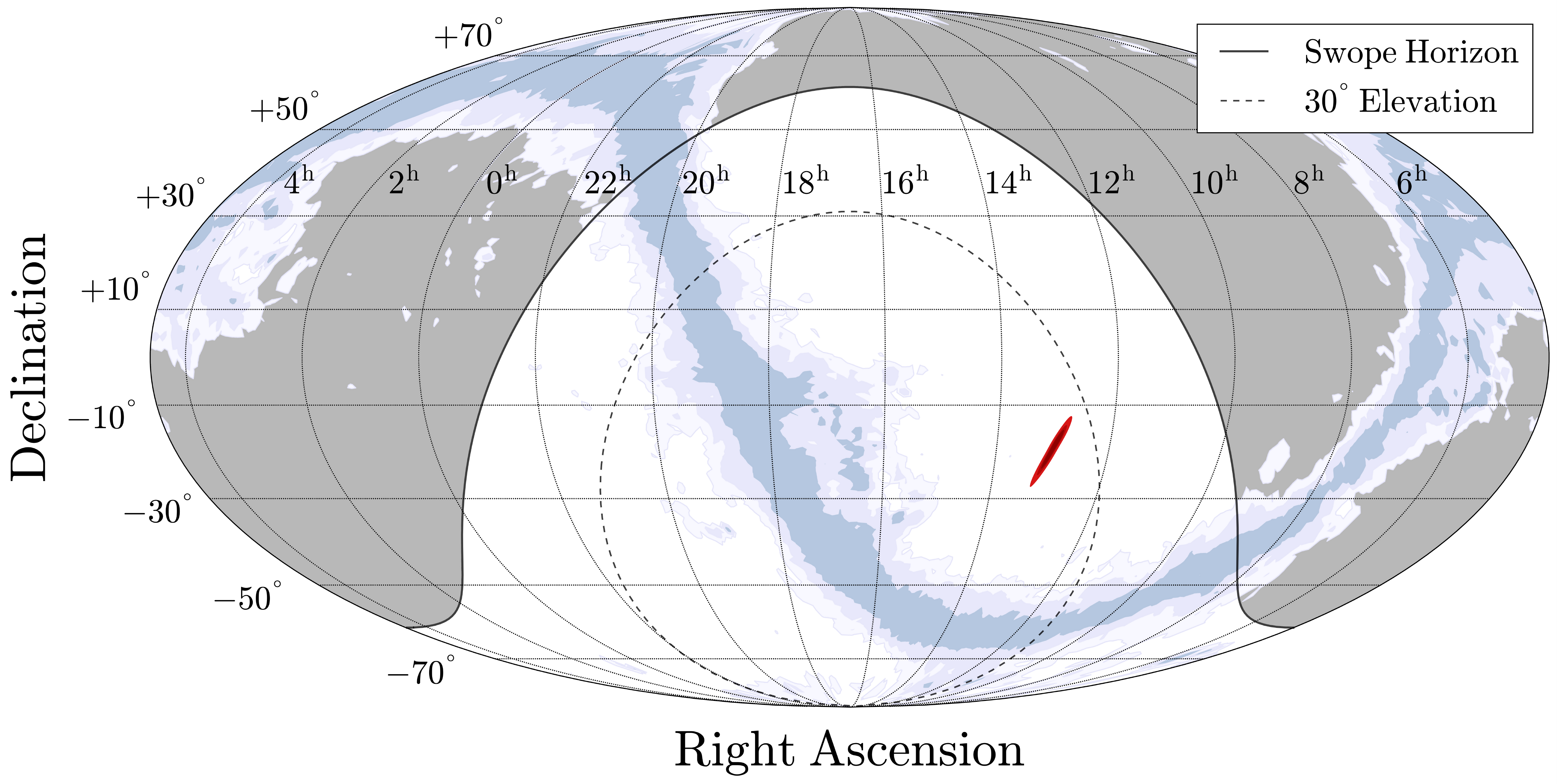}
 \caption{{\bf Gravitational-wave localization of GW170817.}  The outer edge of the red region represents the 90th-percentile confidence region as extracted from the revised {\tt BAYESTAR} probability map. Also shown is the Milky Way in blue for context, with the outermost blue contour corresponding to $V$-band extinction $A_{V} = 0.5$~mag \cite{Schlafly11}.  The thick solid line represents the horizon as seen from the Swope telescope on 2017 August 17 at 23:33 UT, the time we observed SSS17a.  The dotted line represents an elevation above the horizon of 30$^{\circ}$ (corresponding to an airmass of 2.0).
 }
 \end{center}
 \end{figure}

% \noindent
% 1 {\bf Gravitational-wave localization of GW170817.}  The outer edge of the red region represents the 90th-percentile confidence region as extracted from the revised {\tt BAYESTAR} probability map. Also shown is the Milky Way in blue for context, with the outermost blue contour corresponding to $V$-band extinction $A_{V} = 0.5$~mag \cite{Schlafly11}.  The thick solid line represents the horizon as seen from the Swope telescope on 2017 August 17 at 23:33 UT, the time we observed SSS17a.  The dotted line represents an elevation above the horizon of 30$^{\circ}$ (corresponding to an airmass of 2.0).

\clearpage

 \begin{figure}
 \begin{center}
 \includegraphics[angle=0,width=6in]{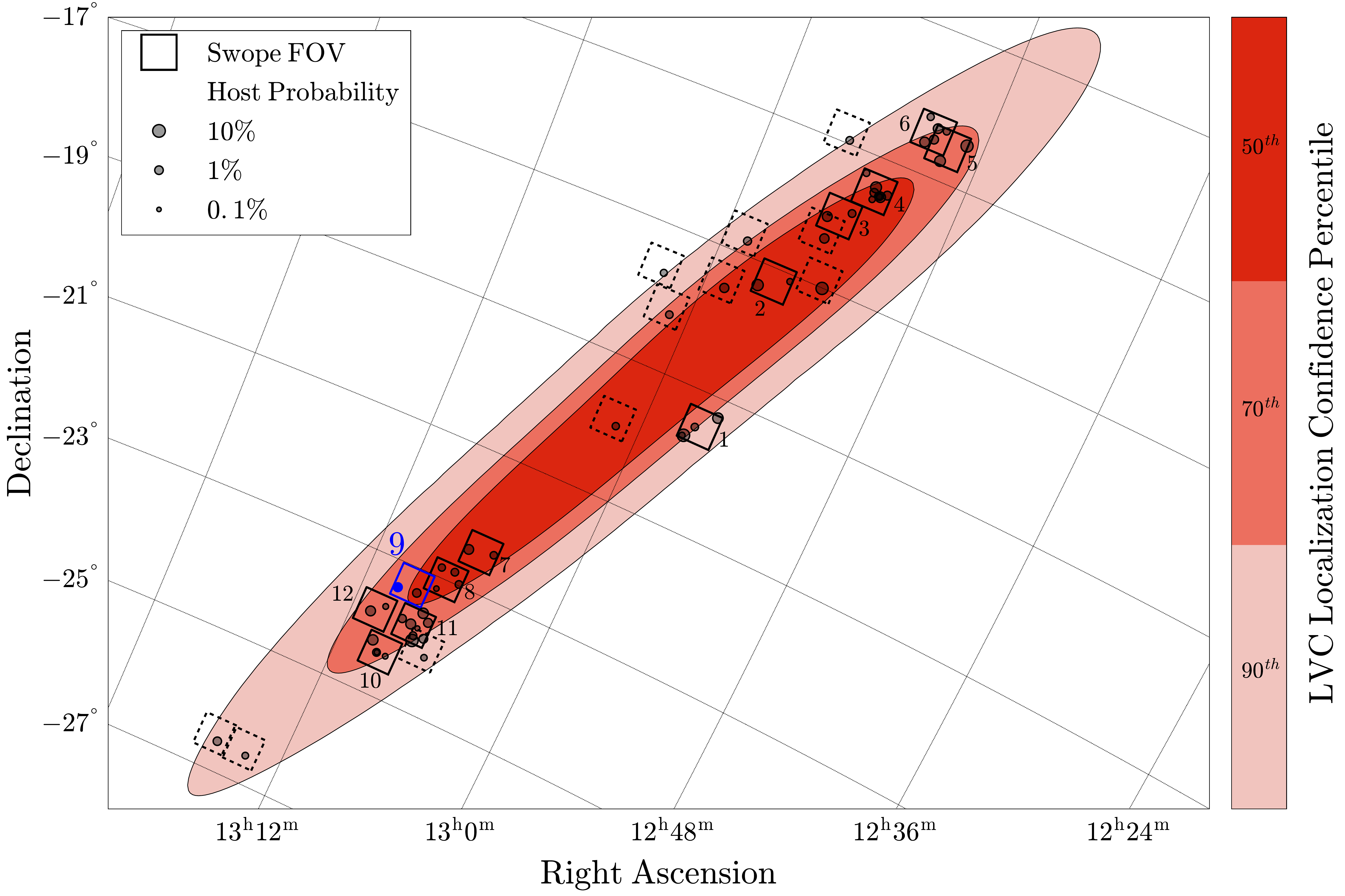}
 \caption{{\bf Sky region covering the 90th-percentile confidence region for the location of GW170817.}  The 50th, 70th, and 90th-percentile contours are shown, with contours extracted from the same probability map as Fig.~1.  Grey circles represent the locations of galaxies in our galaxy catalog and observed by the Swope telescope on 2017 August 17-18 to search for the EM counterpart to GW170817.  The size of the circle indicates the probability of a particular galaxy being the host galaxy for GW170817.  The square regions represent individual Swope pointings with the solid squares specifically chosen to contain multiple galaxies (and labeled in the order that they were observed) and the dotted squares being pointings which contained individual galaxies.  The blue square labeled '9' contains NGC~4993, whose location is marked by the blue circle, and SSS17a.}
 \end{center}
 \end{figure}

% \noindent
% 2 {\bf Sky region covering the 90th-percentile confidence region for the location of GW170817.}  The 50th, 70th, and 90th-percentile contours are shown, with contours extracted from the same probability map as Fig.~1.  Grey circles represent the locations of galaxies in our galaxy catalog and observed by the Swope telescope on 2017 August 17-18 to search for the EM counterpart to GW170817.  The size of the circle indicates the probability of a particular galaxy being the host galaxy for GW170817.  The square regions represent individual Swope pointings with the solid squares specifically chosen to contain multiple galaxies (and labeled in the order that they were observed) and the dotted squares being pointings which contained individual galaxies.  The blue square labeled '9' contains NGC~4993, whose location is marked by the blue circle, and SSS17a.

\clearpage

 \begin{figure}
 \begin{center}
 \includegraphics[angle=0,width=6in]{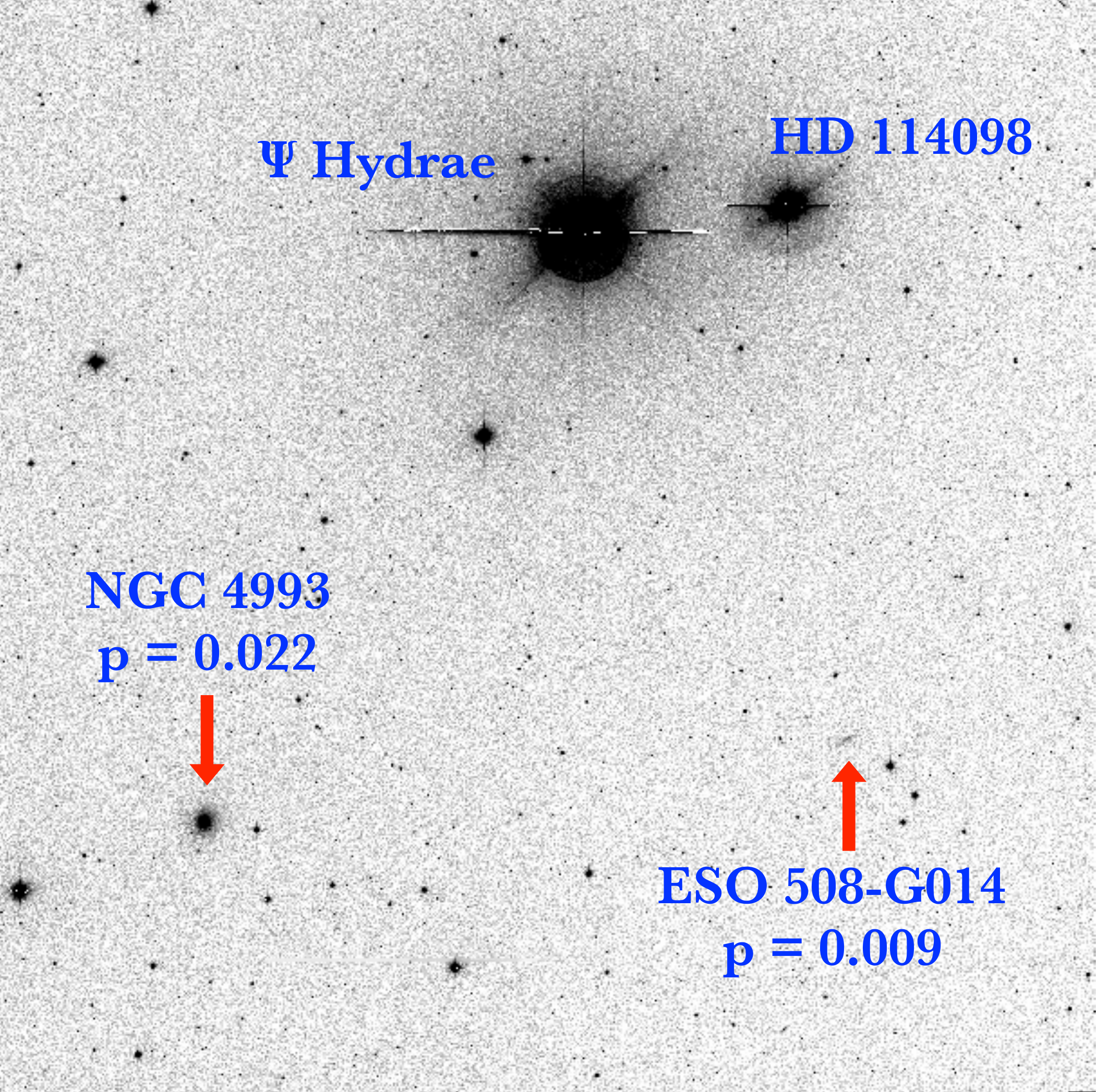}
 \caption{{\bf Full-field Swope telescope $i$-band image containing NGC~4993 (Field 9 in Fig.~2).}  The bright stars $\Psi$~Hydrae and HD~114098 are labeled.  The galaxies NGC~4993 and ESO~508-G014, which had probabilities of hosting GW170817 of 0.022 and 0.009 (see Supplemental Information), respectively, are labeled and marked with red arrows.}
 \end{center}
 \end{figure}

% \noindent
% 3 {\bf Full-field Swope telescope $i$-band image containing NGC~4993 (Field 9 in Fig.~2).}  The bright stars $\Psi$~Hydrae and HD~114098 are labeled.  The galaxies NGC~4993 and ESO~508-G014, which had probabilities of hosting GW170817 of 0.022 and 0.009 (see Supplemental Information), respectively, are labeled and marked with red arrows.

\clearpage

 \begin{figure}
 \begin{center}
 \includegraphics[angle=0,width=6in]{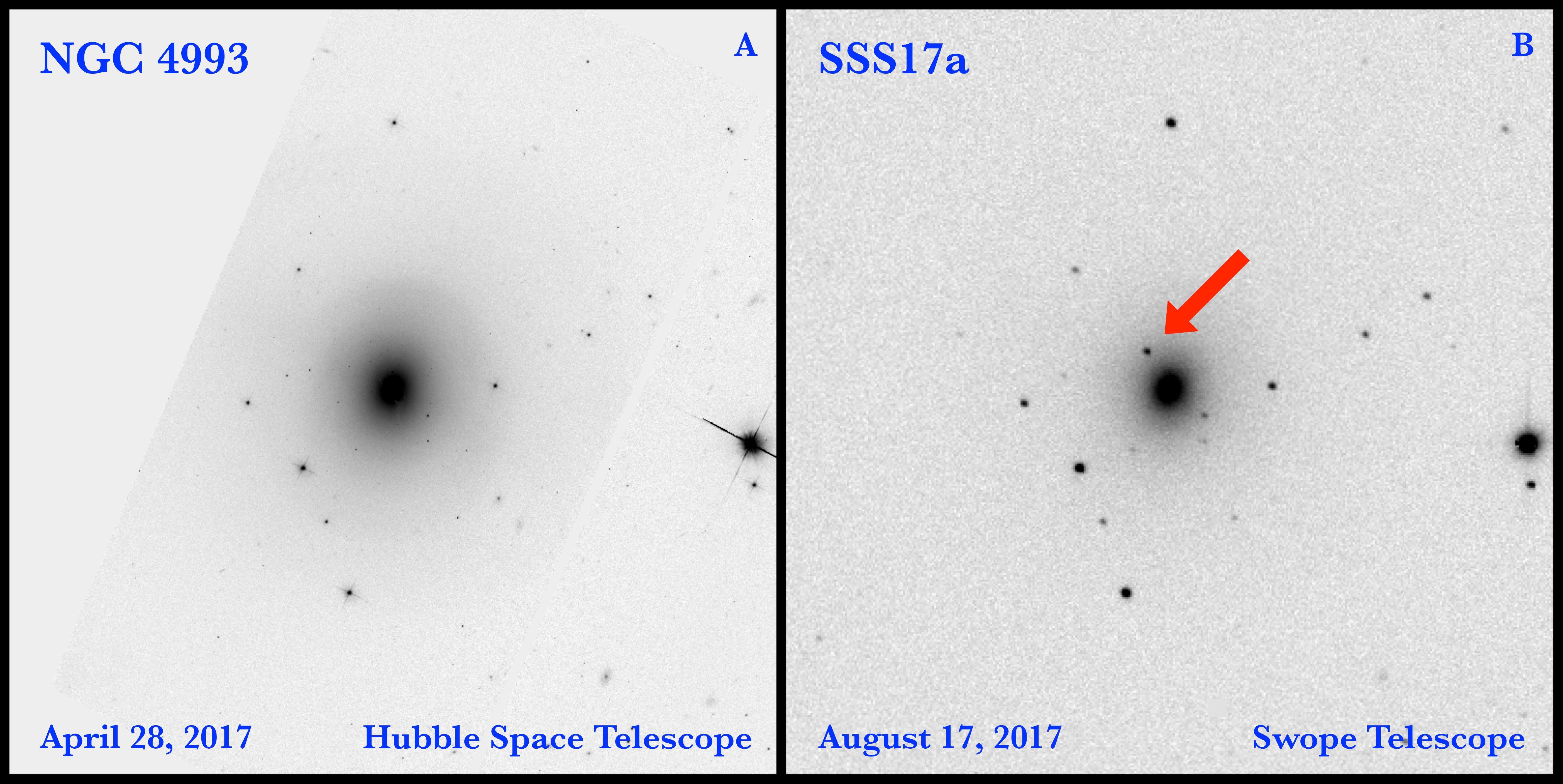}
 \caption{{\bf 3\arcmin\ $\times$ 3\arcmin\ images centered on NGC~4993 
 with North up and East left.}  {\it Panel A}: {\it Hubble Space Telescope} F606W-band (broad $V$) image from 4~months before the GW trigger \cite{GCN21536,Pan17}.  {\it Panel B}: Swope image of SSS17a.  The $i$-band image was obtained on 2017 August 17 at 23:33 UT by the Swope telescope at Las Campanas Observatory.  SSS17a is marked with the red arrow.  No object is present in the {\it Hubble} image at the position of SSS17a \cite{GCN21536,Pan17}.}
 \end{center}
 \end{figure}

% \noindent
% 4 {\bf 3\arcmin\ $\times$ 3\arcmin\ images centered on NGC~4993 
% with North up and East left.}  {\it Panel A}: {\it Hubble Space Telescope} F606W-band (broad $V$) image from 4~months before the GW trigger \cite{GCN21536,Pan17}.  {\it Panel B}: Swope image of SSS17a.  The $i$-band image was obtained on 2017 August 17 at 23:33 UT by the Swope telescope at Las Campanas Observatory.  SSS17a is marked with the red arrow.  No object is present in the {\it Hubble} image at the position of SSS17a \cite{GCN21536,Pan17}.

\clearpage

%Here you should list the contents of your Supplementary Materials -- below is an example. 
%You should include a list of Supplementary figures, Tables, and any references that appear only in the SM. 
%Note that the reference numbering continues from the main text to the SM.
% In the example below, Refs. 4-10 were cited only in the SM.     

\noindent
Materials and Methods\\
Supplementary Text\\
Figs.\ S1 to S6\\
Tables S1 to S2\\
References \textit{(36-65)}

\clearpage

\newpage
\section*{Materials and Methods}
\setcounter{figure}{0}    
\renewcommand{\thefigure}{S\arabic{figure}}
\renewcommand\thesubsection{S\arabic{subsection}:}
\renewcommand\thetable{S\arabic{table}}
\renewcommand\theequation{S\arabic{equation}}

\subsection{Galaxy Prioritization and Scheduling Algorithms}
To improve the chances of detecting an EM counterpart to a GW source, we generate a prioritized list of targets, which we then attempt to combine into as few images as possible, and then produce an observing schedule to be executed by an on-site observer.

We start with a catalog of nearby galaxies specifically created for the purpose of targeted-galaxy GW searches \cite{White11}.  The catalog contains 53,161 galaxies, is nearly complete to 40~Mpc (in terms of stellar mass), and still has high completeness to 100~Mpc.  In addition to positions and distances, the catalog includes $B$-band magnitudes, from which we calculate a $B$-band luminosity that is a proxy for some combination of total stellar mass (more massive galaxies are more luminous) and star-formation rate (SFR; bluer galaxies are more vigorously forming stars).  Both are useful for EM searches of GW sources since more massive galaxies have more BNS systems and the theoretical merger rate depends on the SFR \cite{Phinney91,Belczynski02}.

Our algorithm depends on the three-dimensional localization probability maps produced by LVC.  In the case of GW170817, we used the revised {\tt BAYESTAR} map \cite{GCN21513,Singer16}.  This map contains a two-dimensional (plane of the sky) probability and a distance (and distance uncertainty) for pixels in a Hierarchical Equal Area isoLatitude
Pixelization (HEALPix) map.  Within this map, there were 49 galaxies
in the catalog and in the 90th-percentile localization volume.  Since GW170817 had an associated SGRB \cite{GCN21506,GCN21507}, which might imply a face-on system and thus a larger distance \cite{GCN21513}, we assumed a single distance for every pixel of the probability map with an inflated uncertainty, $D = 43 \pm 12$~Mpc.

For each galaxy in the catalog, we calculate a $B$-band luminosity proxy, $\tilde{L}_{B}$ (uncorrected for Milky Way extinction),
\begin{equation}
  \tilde{L}_{B} = D_{\rm gal}^{2} 10^{-0.4 m_{B}},
\end{equation}
where $D_{\rm gal}$ is the catalog distance for a galaxy and $m_{B}$ is its $B$-band magnitude.

Combining the two-dimensional location, distance, and galaxy luminosity, we obtain our final probability:
\begin{equation}
  P_{\rm gal} = k^{-1} \times \tilde{L}_{B} \times P_{\rm 2D} \times \left ( 1 - \rm{erf} \left ( \frac{| D_{\rm gal} - D_{\rm LVC} |}{\sigma_{D,{\rm ~gal}}^{2} + \sigma_{D, {\rm ~LVC}}^{2}} \right ) \right ),
\end{equation}
where $P_{\rm gal}$ is the probability that a particular galaxy hosts the GW event, $k$ is a normalization factor such that all probabilities sum to 1, $P_{\rm 2D}$ is the two-dimensional probability for a particular galaxy, $D_{\rm gal}$ and $\sigma_{D,{\rm ~gal}}$ are the distance and distance uncertainty for the galaxy, and $D_{\rm LVC}$ and $\sigma_{D,{\rm ~LVC}}$ are the distance and distance uncertainty for the GW source.

We then attempted to schedule the 1000 highest-probability galaxies for observations with the Swope telescope, the Magellan Clay telescope with the LDSS-3 imaging spectrograph \cite{LDSS} and the Magellan Baade telescope with the FourStar near-infrared camera \cite{Persson13}.  Using all three telescopes for the search enabled us to cover a range of optical and near-IR wavelengths over which a counterpart could be detected.  Our scheduling algorithm takes an input of object positions, using the computed probabilities as priorities, and takes into account observational constraints such as the total observing time for each object (exposure plus overhead) to produce a schedule.  The algorithm maximizes a merit function based on a total net priority that includes the object's observing constraints in addition to its computed probability, and it attempts to place the highest net-priority targets at each target's lowest airmass.  For the GW170817 search, all exposure times were identical and every target was setting, which reduced the algorithm to scheduling the highest-probability unobserved targets above our airmass limit of 3 at any given time.

Using our scheduling algorithm, we were able to schedule $\sim 100$ galaxies with non-negligible probability, indicating roughly the number of observations we could perform before all high-probability galaxies set below our airmass limit.  We then examined the positions of the 100 highest-probability galaxies to determine if multiple galaxies could be observed simultaneously.  We visually examined the galaxy positions and were able to group 46 galaxies into 12 separate telescope pointings, improving our efficiency by a factor of 3.8 for the grouped galaxies, and 52\% for the 100 highest-priority galaxies.

We added special targets corresponding to the positions of the multi-galaxy fields and assigned them a high (and equal) priority to guarantee that they would be observed.  By including the multi-galaxy fields, we freed up time to observe 34 additional high-probability galaxies.  After swapping galaxies for multi-galaxy fields covering those same galaxies in our target list, we recomputed an observing schedule.  We did not attempt to further optimize the pointings.

Table~S2 contains a list of observed galaxies, their probabilities, and the order of observation for each telescope.

\begin{longtable}{@{}lllcccc}
\caption{Observation Schedule}\\
\hline
\hline
Galaxy & R.A. & Decl. & Probability & \multicolumn{3}{c}{Observation Number}\\   &   &   &   & Swope & LDSS-3 & FourStar \\ \hline
NGC 4830      & 12:57:27.9      & -19:41:28      & 0.086207      & 1      & 2      & 1\\
NGC 4970      & 13:07:33.7      & -24:00:31      & 0.083333      & 11      & 3      & 9\\
NGC 4763      & 12:53:27.2      & -17:00:18      & 0.077519      & 13      & 4      & 2\\
IC 3799      & 12:48:59.7      & -14:23:56      & 0.073529      & 5      & 5      & 3\\
PGC 044234      & 12:57:00.5      & -17:19:13      & 0.044248      & 2      & 6      & 4\\
NGC 4756      & 12:52:52.6      & -15:24:48      & 0.037037      & 4      & 7      & 5\\
PGC 043424      & 12:50:04.7      & -14:44:00      & 0.034014      & 5      & 8      & 6\\
ESO 575-G029      & 12:55:59.7      & -19:16:07      & 0.028818      & 1      & 9      & \\
ESO 508-G010      & 13:07:37.8      & -23:34:43      & 0.027855      & 11      & 10      & \\
PGC 043664      & 12:52:25.6      & -15:31:02      & 0.026316      & 4      & 1      & 7\\
ESO 508-G019      & 13:09:51.7      & -24:14:22      & 0.025773      & 10      & 11      & \\
\textbf{NGC 4993}      & \textbf{13:09:47.7}      & \textbf{-23:23:01}      & \textbf{0.021463}      & \textbf{9}      & \textbf{12}      & \textbf{11}\\
IC 4197      & 13:08:04.3      & -23:47:49      & 0.021368      & 11      &       & \\
ESO 508-G024      & 13:10:45.9      & -23:51:56      & 0.020243      & 12      &       & \\
PGC 043966      & 12:54:49.5      & -16:03:08      & 0.019531      & 3      &       & \\
IC 3831      & 12:51:18.6      & -14:34:24      & 0.019531      & 6      &       & \\
ESO 575-G055      & 13:06:39.9      & -22:27:20      & 0.018762      & 7      &       & \\
NGC 4724      & 12:50:53.6      & -14:19:54      & 0.018116      & 5, 6      &       & \\
PGC 043908      & 12:54:28.9      & -16:21:03      & 0.015152      & 14      &       & \\
PGC 044500      & 12:58:45.6      & -17:32:35      & 0.013812      & 15      &       & \\
IC 3827      & 12:50:52.2      & -14:29:30      & 0.012453      & 5, 6      &       & \\
NGC 4968      & 13:07:05.8      & -23:40:37      & 0.012225      & 11      &       & \\
PGC 043625      & 12:52:05.4      & -15:27:29      & 0.011403      & 4      &       & \\
IC 4180      & 13:06:56.5      & -23:55:01      & 0.010776      & 11      &       & 10\\
PGC 043720      & 12:52:51.1      & -15:29:29      & 0.009615      & 4      & 1      & 7\\
ESO 508-G014      & 13:08:32.3      & -23:20:49      & 0.009174      & 9      &       & \\
ESO 508-G033      & 13:16:23.3      & -26:33:42      & 0.009091      & 16      &       & \\
PGC 797164      & 13:08:42.5      & -23:46:32      & 0.007634      & 11      &       & \\
PGC 044478      & 12:58:34.4      & -16:48:16      & 0.007246      & 17      &       & \\
PGC 043662      & 12:52:29.4      & -15:29:57      & 0.006536      & 4      & 1      & 7\\
ESO 508-G015      & 13:09:18.9      & -24:23:05      & 0.006211      & 10      &       & 8\\
PGC 043808      & 12:53:33.9      & -15:52:44      & 0.006061      & 3      &       & \\
ESO 508-G004      & 13:06:52.6      & -22:50:29      & 0.005952      & 8      &       & \\
PGC 183552      & 13:07:37.7      & -23:56:17      & 0.005848      & 11      & 3      & 9, 10\\
PGC 044021      & 12:55:19.3      & -14:57:00      & 0.005780      & 18      &       & \\
PGC 169673      & 13:06:19.4      & -22:58:48      & 0.005747      & 8      &       & \\
IC 0829      & 12:52:33.0      & -15:31:00      & 0.005682      & 4      & 1      & 7\\
ESO 575-G047      & 13:01:09.2      & -18:11:51      & 0.005556      & 19      &       & \\
ESO 575-G035      & 12:57:02.7      & -19:31:05      & 0.005319      & 1      &       & \\
ESO 575-G053      & 13:05:04.9      & -22:23:01      & 0.005102      & 7      &       & \\
ESO 575-G048      & 13:01:26.8      & -19:57:52      & 0.004762      & 20      &       & \\
ESO 508-G011      & 13:07:44.9      & -22:51:28      & 0.004608      & 8      &       & \\
NGC 4726      & 12:51:32.4      & -14:13:16      & 0.004525      & 6      &       & \\
IC 3822      & 12:50:22.7      & -14:19:18      & 0.004425      & 5, 6      &       & \\
PGC 045006      & 13:02:25.9      & -17:40:47      & 0.004149      & 21      &       & \\
PGC 043823      & 12:53:42.3      & -15:16:56      & 0.003401      & 4      &       & \\
PGC 046026      & 13:14:17.7      & -26:34:58      & 0.003390      & 22      &       & \\
PGC 043663      & 12:52:27.4      & -15:31:07      & 0.003195      & 4      & 1      & 7\\
ESO 508-G003      & 13:06:24.0      & -24:09:50      & 0.002950      & 23      &       & \\
PGC 043711      & 12:52:48.9      & -15:35:21      & 0.002695      & 4      &       & 7\\
PGC 044312      & 12:57:32.7      & -19:42:01      & 0.002660      & 1      & 2      & 1\\
PGC 044023      & 12:55:20.4      & -17:05:46      & 0.002639      & 2      &       & \\
NGC 5124      & 13:24:50.3      & -30:18:27      & 0.002340      & 24      &       & \\
NGC 5051      & 13:16:20.1      & -28:17:09      & 0.002290      & 25      &       & \\
ESO 508-G020      & 13:09:59.8      & -23:42:51      & 0.002045      & 12      &       & \\
PGC 045565      & 13:08:42.0      & -24:22:58      & 0.001905      & 10      &       & 8\\
PGC 803966      & 13:07:30.9      & -23:10:14      & 0.001736      & 8      &       & \\
NGC 5078      & 13:19:50.1      & -27:24:36      & 0.001630      & 26      &       & \\
ESO 444-G012      & 13:20:50.2      & -29:28:46      & 0.001580      & 27      &       & \\
ESO 444-G026      & 13:24:29.0      & -30:25:54      & 0.001560      & 28      &       & \\
6dF J1309178-242256      & 13:09:17.7      & -24:22:55      & 0.001520      & 10      &       & 8\\
Abell 1664-11      & 13:07:34.1      & -23:48:54      & 0.001420      & 11      &       & 10\\
NGC 5114      & 13:24:01.7      & -32:20:38      & 0.001300      & 29      &       & \\
NGC 5135      & 13:25:44.0      & -29:50:00      & 0.001260      & 30      &       & \\
ESO 444-G021      & 13:23:30.6      & -30:06:51      & 0.001250      & 31      &       & \\
NGC 5048      & 13:16:08.4      & -28:24:37      & 0.001160      & 32      &       & \\
NGC 5193      & 13:31:53.5      & -33:14:03      & 0.001100      & 33      &       & \\
ESO 383-G005      & 13:29:23.6      & -34:16:16      & 0.000937      & 34      &       & \\
NGC 5140      & 13:26:21.7      & -33:52:07      & 0.000914      & 35      &       & \\
2MASX J13245297-3020059      & 13:24:53.0      & -30:20:04      & 0.000429      & 36      &       & \\
ESO 444-G019      & 13:23:06.3      & -32:14:41      & 0.000328      & 37      &       & \\
IC 4296      & 13:36:39.1      & -33:57:57      & 0.000290      & 38      &       & \\
ESO 221-G035      & 14:16:04.4      & -52:36:30      & 0.000188      & 39      &       & 22\\
ESO 221-G030      & 14:10:41.1      & -52:11:02      & 0.000171      & 40      &       & 23\\
ESO 175-G002      & 14:08:36.0      & -53:21:10      & 0.000149      & 41      &       & \\
ESO 383-G027      & 13:35:04.9      & -35:16:08      & 0.000141      & 42      &       & \\
ESO 324-G033      & 13:32:27.3      & -38:10:05      & 0.000103      & 43      &       & 13\\
ESO 383-G047      & 13:37:50.6      & -36:03:00      & 0.000077      & 44      &       & \\
ESO 324-G044      & 13:38:06.2      & -39:50:25      & 0.000066      & 45      &       & 12\\
ESO 221-G020      & 13:58:23.1      & -48:28:34      & 0.000060      & 46      &       & \\
PGC 141857      & 14:10:33.5      & -52:19:02      & 0.000055      & 47      &       & 14\\
PGC 2800412      & 14:17:10.0      & -55:37:11      & 0.000051      & 48      &       & 15\\
PGC 141859      & 14:20:23.5      & -55:04:07      & 0.000051      & 49      &       & 16\\
PGC 166335      & 14:16:02.0      & -53:42:59      & 0.000043      & 50      &       & 18\\
NGC 5365A      & 13:56:39.5      & -44:00:32      & 0.000038      & 51      &       & 17\\
PGC 463082      & 14:03:29.3      & -50:46:37      & 0.000037      & 52      &       & 19\\
PGC 166323      & 14:04:34.1      & -52:41:49      & 0.000035      & 53      &       & 20\\
ESO 175-G005      & 14:17:47.0      & -52:49:54      & 0.000033      & 54      &       & 21\\
\hline
\label{tab:Observations}
\end{longtable}

While NGC~4993 was included in a multi-galaxy field, skipping this step would have delayed our imaging of NGC~4993 by only $\sim$6~minutes.  That is, NGC~4993 was in the ninth image obtained, but was the twelfth-highest probability galaxy.

We applied the same algorithms to produce observing schedules for Magellan Clay/LDSS-3 and Magellan Baade/FourStar.  Maps displaying the locations of galaxies observed by these telescopes are shown in Fig.~S1 \& S2, respectively.

\begin{figure}
\begin{center}
\includegraphics[angle=0,width=6in]{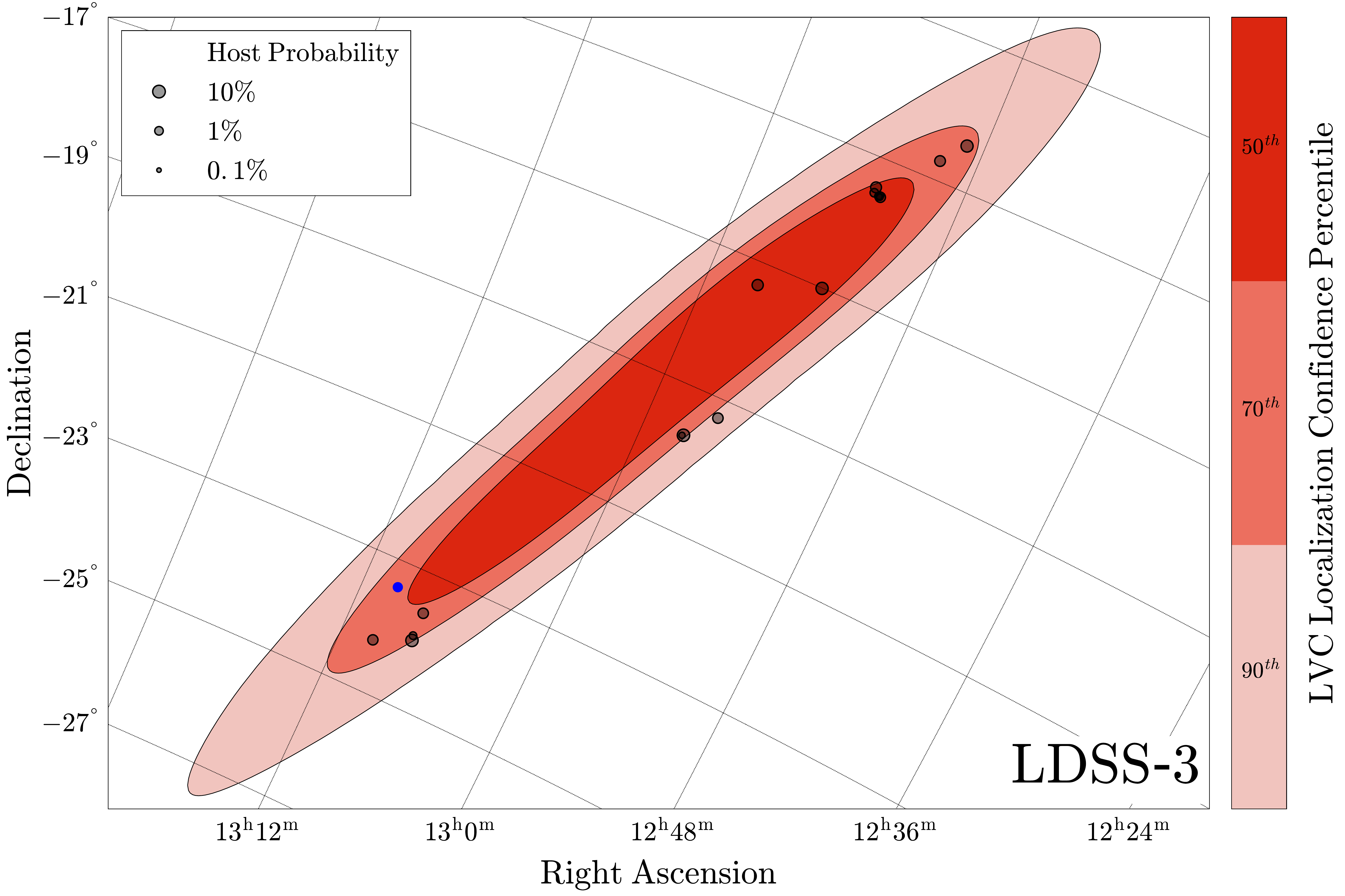}
\caption{Same as Fig.~2, except displaying the galaxies observed by Magellan Clay/LDSS-3. NGC~4993 is marked by the blue circle.}
\end{center}
\end{figure}

\begin{figure}
\begin{center}
\includegraphics[angle=0,width=6in]{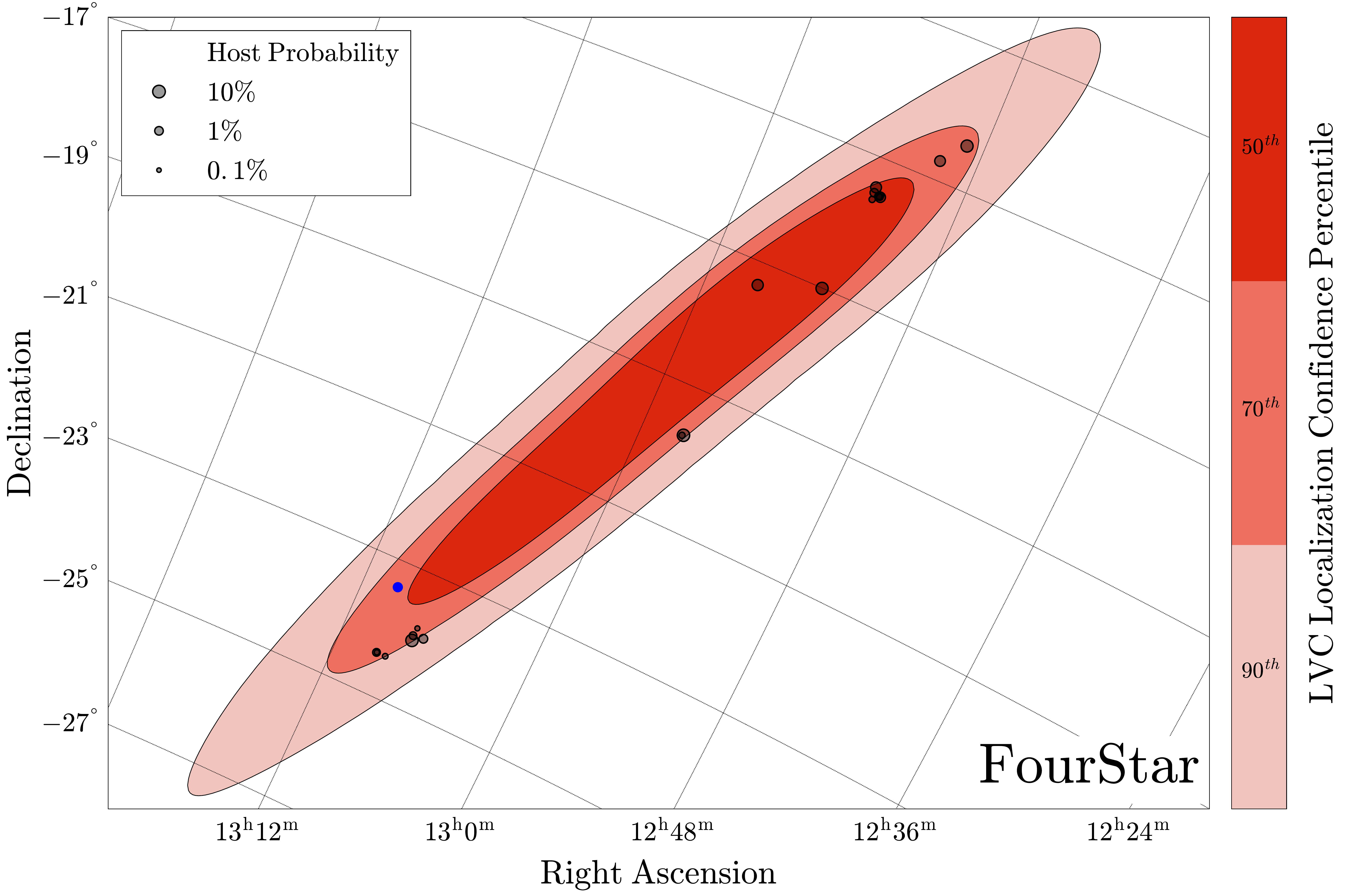}
\caption{Same as Fig.~2, except displaying the galaxies observed by Magellan Baade/FourStar. NGC~4993 is marked by the blue circle.}
\end{center}
\end{figure}

\subsection{Initial Transient Search}
In preparing to search the LIGO/Virgo localization region for potential optical counterparts, we obtained template images of our galaxy fields from the National Optical Astronomy Observatory (NOAO) public archive \cite{NOAO}.  Our preference was to obtain deep template images covering as many galaxies in our sample over as wide an area as possible and in a similar photometric band to the Swope $i$ filter.  Therefore, we searched the NOAO archive for Dark Energy Camera (DECam) \cite{Diehl08} $r$- and $i$-band images covering as large a region as many of our galaxy fields as possible.

%% Re: ds9. Is there an instrument description paper you can cite?
With the DECam reference images, we immediately identified the 100 highest-probability galaxies using {\tt ds9} \cite{Joye03} region files.  As each Swope field was observed, we reduced the images using a reduction pipeline (described below).  We then loaded each file into {\tt ds9} to visually inspect the galaxies associated with each region.

Transient discovery by visual inspection is heavily biased toward bright, isolated sources, and so we used multiple kinds of scaling to inspect sources around each galaxy.  This process included inspection of faint sources in the outskirts of each galaxy using typical linear scaling and sources deeply embedded in each galaxy with significantly higher clip values in order to resolve faint, nuclear sources.  As we inspected each image, we blinked them rapidly with the DECam images in order to identify any differences between the images.  Using this method, we were able to visually inspect a single image every $\sim$1~minute, which was appreciably faster than the total time allocated to expose on each field, read out the instrument, and slew to each successive field.  However, accounting for image reductions, we visually inspected the initial images roughly $20$~minutes after they were obtained.  

The image containing SSS17a was the ninth image obtained, reduced, and inspected.  As we demonstrate in Figs.~3 \& 4, this field contained only two galaxies and the transient source was immediately apparent upon comparison with template imaging.  

After the identification of this source at 23:59 UT (Fig.~S6), we queried the Minor Planet Center \cite{MPC} and Transient Name Server \cite{TNS} databases to confirm that SSS17a was not a known asteroid or SN.  Upon confirmation that SSS17a was an unknown source, our priorities immediately changed to characterization through follow-up observations. After additional observations, we continued our search to observe the remaining galaxies in our list.  The fields observed are presented in Figure~S3.

\begin{figure}
\begin{center}
\includegraphics[angle=0,width=3in]{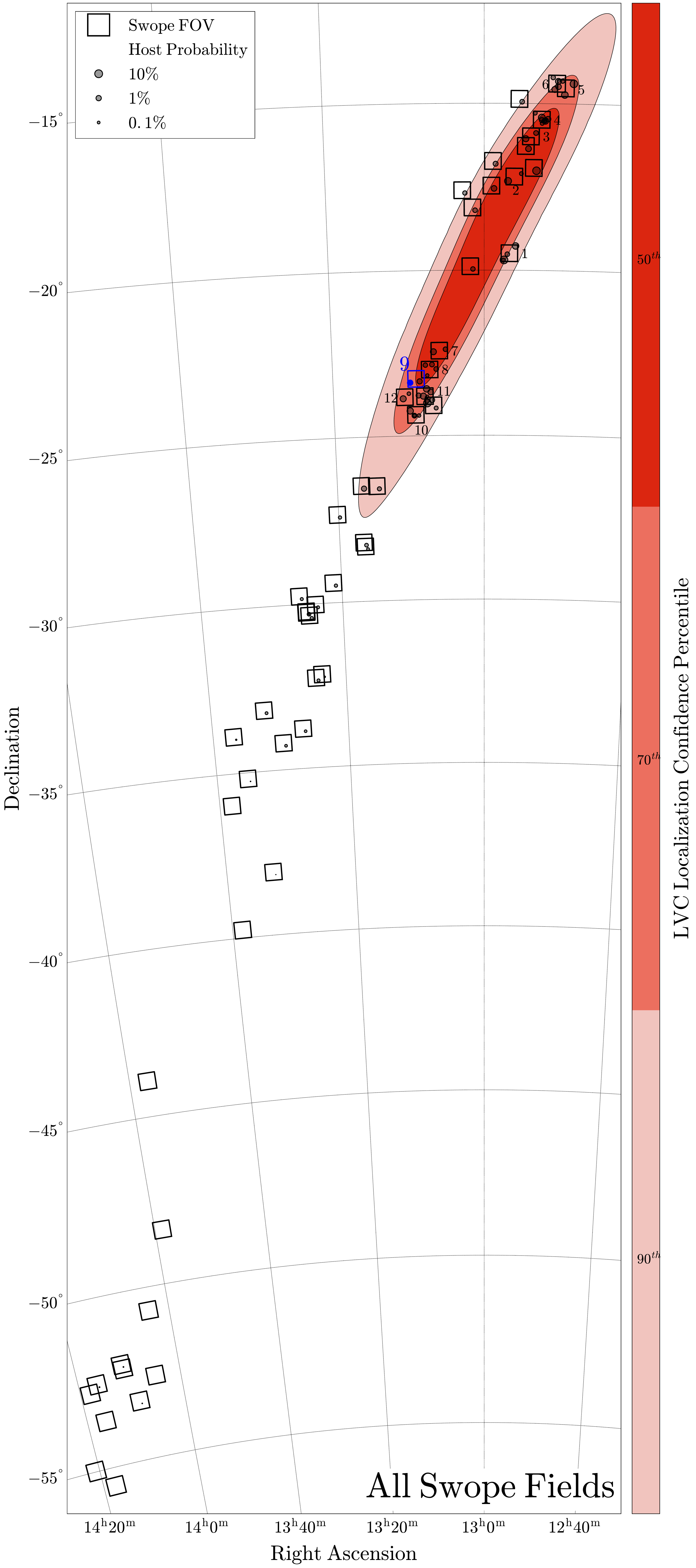}
\caption{Same as Fig.~2, except displaying all galaxies and fields observed by Swope. NGC~4993 is marked by the blue circle.}
\end{center}
\end{figure}

\subsection{Astrometry}
We determined the location of SSS17a from the centroid of our point-spread function (PSF) fit in the discovery $i$ band image and the world-coordinate system (WCS) solution derived for that image.  The astrometric uncertainty associated with the location of SSS17a is the $1$-$\sigma$ uncertainty on the PSF centroid and on the WCS solution added in quadrature.  Our total astrometric uncertainty for SSS17a, which was detected at $17.476$~mag in a relatively bright galaxy, was about $0.23\arcsec$.  

\subsection{Swope Photometry}
Subsequent to discovery, we imaged the site of SSS17a in $BV\!gri$ bands with the Swope telescope from 2017 August 17-24, at which point it became too faint to detect in $30$~minute $i$ band images.  After that time, we imaged the same field with long exposure times ($20$--$60$~minute) in order to construct deep templates for difference imaging.

We performed standard reductions on all of our Swope imaging using {\tt photpipe} \cite{Rest05:photpipe}, a well-tested pipeline used in the Swope Supernova Survey and several major time-domain surveys \cite[e.g., Pan-STARRS1]{Rest14}.  We used {\tt photpipe} to correct the Swope images for bias, flat-fielding, cross-talk, and overscan.  We performed astrometric calibration and corrections for geometric distortion in {\tt photpipe} by resampling each image onto a corrected grid with {\tt SWarp} \cite{Bertin02}, then applied a WCS derived from the locations of stars in the 2MASS Point Source Catalog \cite{Skrutskie06}.  We used {\tt photpipe} to perform difference imaging with {\tt hotpants} \cite{Alard00,Becker15}, which uses a spatially varying convolution kernel to match the image and template PSF before image subtraction.  As an example, we display the discovery image after image subtraction in Fig.~S4.  Finally, we performed photometry using {\tt DoPhot}, which is optimized for PSF photometry on the difference images \cite{Schechter93}.  Our $BV\!gri$ photometry was calibrated using Pan-STARRS1 (PS1) standard stars \cite{Chambers16,Flewelling16,Magnier16,Waters16} observed in the same field as SSS17a and transformed into the Swope natural system using Supercal transformations \cite{Scolnic15}.

The Swope photometry is presented in Fig.~S5 and listed in Table~S1.

\begin{figure}
\begin{center}
\includegraphics[angle=0,width=6in]{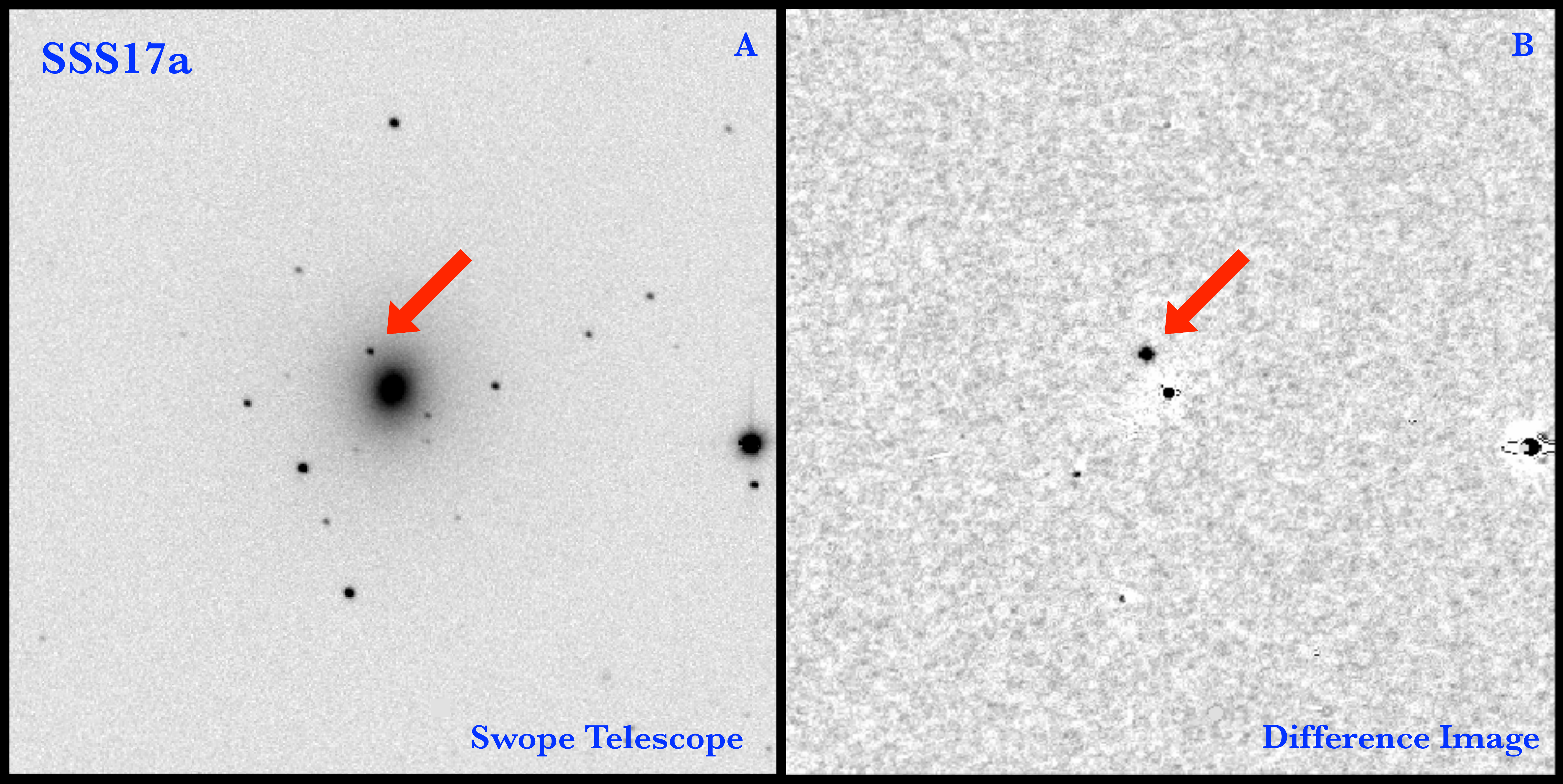}
\caption{3\arcmin\ $\times$ 3\arcmin\ images centered on NGC~4993 
with North up and East left.  {\it Panel A}: Discovery image of SSS17a (same as Fig.~4B).  {\it Panel B}: Difference image of the left panel.  SSS17a is marked with the red arrow.}
\end{center}
\end{figure}

\begin{figure}
\begin{center}
\includegraphics[angle=0,width=4in]{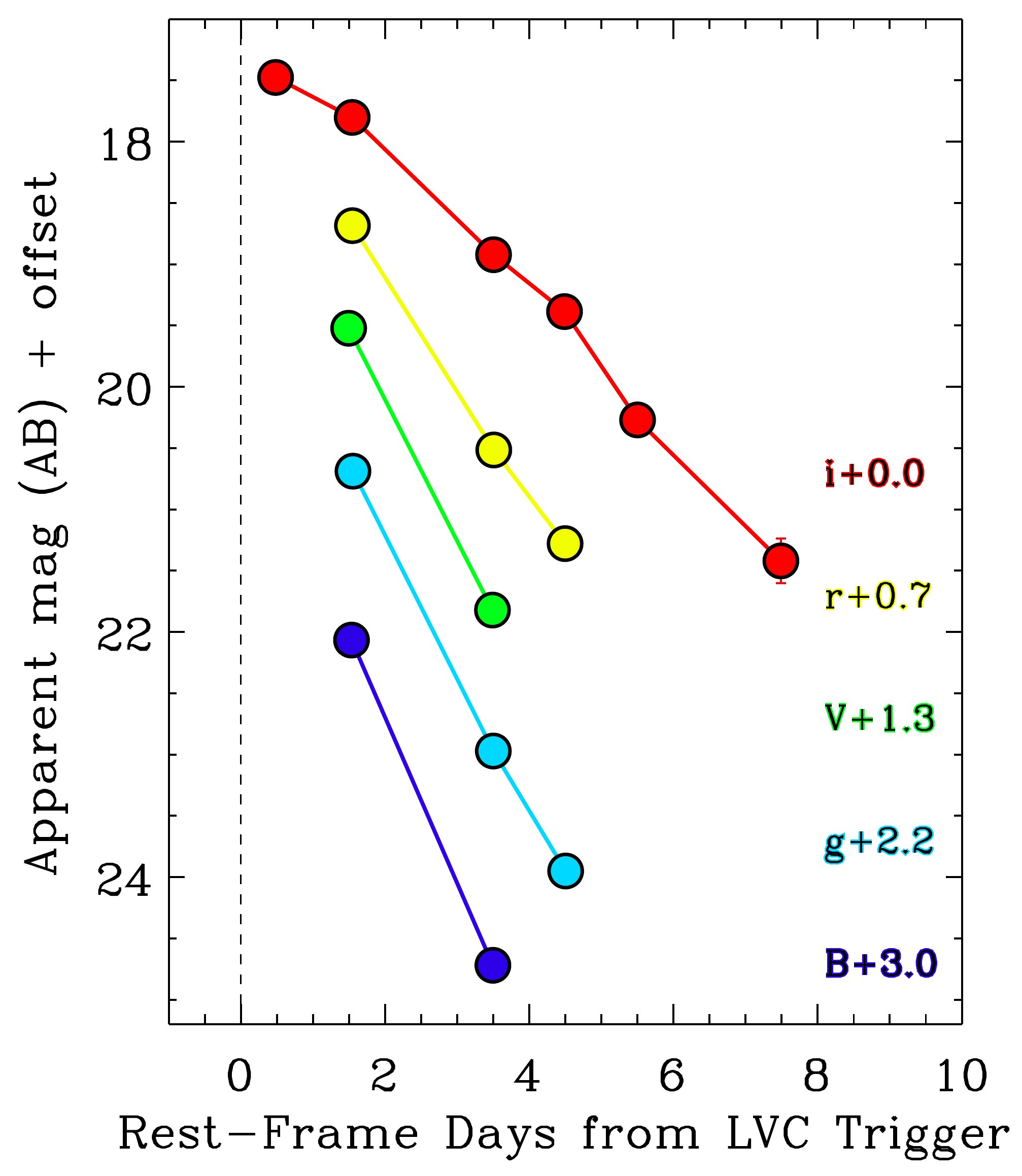}
\caption{$BV\!gri$ light curves of SSS17a.  These data are analyzed in detail in \cite{Drout17,Kilpatrick17,Murguia-Berthier17}.}
\end{center}
\end{figure}

\begin{longtable}{@{}ccc}
\caption{Swope Photometry of SSS17a}\\
\hline\hline
Time since LVC Trigger (d) & Filter & AB Magnitude (Uncertainty) \\ \hline
0.4529 & $i$ & 17.476 (018) \\
1.4663 & $V$ & 18.222 (041) \\
1.5057 & $B$ & 19.066 (037) \\
1.5153 & $i$ & 17.802 (020) \\
1.5187 & $r$ & 17.985 (018) \\
1.5263 & $g$ & 18.488 (124) \\
3.4608 & $V$ & 20.521 (115) \\
3.4666 & $B$ & 21.719 (126) \\
3.4728 & $g$ & 20.771 (049) \\
3.4761 & $i$ & 18.922 (047) \\
3.4791 & $r$ & 19.815 (089) \\
4.4600 & $i$ & 19.386 (045) \\
4.4680 & $r$ & 20.579 (125) \\
4.4759 & $g$ & 21.750 (104) \\
5.4735 & $i$ & 20.270 (116) \\
7.4612 & $i$ & 21.420 (183) \\
\hline
% \begin{tablenotes}
% \item {\bf Note.} Time is in rest-frame days since LVC trigger.  Uncertainties ($1\sigma$) are in millimagnitudes and given in parentheses next to each measurement. $BV$ magnitudes are on the Vega scale and $ugri$ are on the AB scale.
% \end{tablenotes}
\end{longtable}

\subsection{Extensive Transient Search}
From 2017 September 4-7, we obtained follow-up $120$ second $i$-band images of all fields observed with the Swope telescope on 2017 August 17.  We performed difference imaging in the same manner using {\tt photpipe} to search for new optical transients that had either faded since 2017 August 17, or appeared over the 18--20~day interval from the first to second epoch of observation.  Apart from SSS17a, we did not detect any transients in any of our Swope fields.

We calculated a $5$-$\sigma$ limiting magnitude for each Swope image by examining sources with background fluxes within $10\%$ of the median sky background.  We then calculated the signal-to-noise ratio for each of those sources as estimated from the count rate and $1$-$\sigma$ uncertainty in the count rate.  We fit a cubic spline to the signal-to-noise ratio versus magnitude and estimated the magnitude at which the signal-to-noise ratio was equal to $5.0$.  This $5$-$\sigma$ limit was typically around $20.0$~mag in the $60$ second images from 2017 August 17, and $20.8$~mag in the $120$ second images from 2017 September 4-7.

Treating this $5$-$\sigma$ limit as a bolometric magnitude at the distance of NGC~4993, we would expect to detect all sources down to $M = -13.0$ to $-12.2$~mag.  This limiting magnitude rules out the presence of most supernovae, even those observed within hours \cite{Foley12:09ig,Pan15:13dy,Shappee16} or weeks after explosion \cite{Li11:rate2}.  Very low-luminosity transients such as classical novae and intermediate luminosity red transients \cite{Mould90} are faint enough to be missed by our optical survey, but neither of these classes of sources are thought to be strong sources of GW emission.

Our choice of filter was also designed to target EM emission associated with predictions from kilonova models, which are expected to have very high optical opacities and thus very red colors \cite{Li98}.  If the theoretical predictions were incorrect and the intrinsic color of EM counterparts are blue, it is still unlikely that any source would be so blue to be detected in a bluer band (e.g., $g$) and not $i$.  SSS17a is well-fit by kilonova models \cite{Drout17,Kilpatrick17}, but with an added component that is hot ($>10,000$~K) within hours of the LVC trigger and cooled rapidly over several days.

\subsection{1M2H Slack Conversation}
At the time of the GW170817 alert, D.\ Coulter, R.\ Foley, and M.\ Siebert were at the Dark Cosmology Centre in Copenhagen, Denmark, while C.\ Kilpatrick and C.\ Rojas-Bravo were in Santa Cruz, California.  Meanwhile, B.\ Shappee, J.\ Simon, and N.\ Ulloa were at Las Campanas Observatory with M.\ Drout supporting from Pasadena, California.  Both because of the multiple locations and for its speed, we used Slack to communicate.  In Fig.~S6 below, we present our conversation, which includes the timeline of our evolving strategy and discovery of SSS17a.  All times displayed are Pacific Daylight Time.


\begin{thebibliography}{33}

\bibitem{Thorne87}
K.~S. {Thorne}, {\it {Gravitational radiation.}\/} (1987), pp. 330--458.

\bibitem{Abbott16:gw}
B.~P. {Abbott}, {\it et~al.\/}, {\it Phys. Rev. Lett.\/} {\bf 116}, 061102
  (2016).

\bibitem{Abbott16:GW151226}
B.~P. {Abbott}, {\it et~al.\/}, {\it Phys. Rev. Lett.\/} {\bf 116}, 241103
  (2016).

\bibitem{Abbott17}
B.~P. {Abbott}, {\it et~al.\/}, {\it Phys. Rev. Lett.\/} {\bf 118}, 221101
  (2017).

\bibitem{Abbott16:bbh}
B.~P. {Abbott}, {\it et~al.\/}, {\it Phys. Rev. X\/} {\bf 6}, 041015 (2016).

\bibitem{Abbott16:gr}
B.~P. {Abbott}, {\it et~al.\/}, {\it Phys. Rev. Lett.\/} {\bf 116}, 221101
  (2016).

\bibitem{Li98}
L.-X. {Li}, B.~{Paczy{\'n}ski}, {\it \apjBib\/} {\bf 507}, L59 (1998).

\bibitem{Metzger10}
B.~D. {Metzger}, {\it et~al.\/}, {\it \mnrasBib\/} {\bf 406}, 2650 (2010).

\bibitem{Roberts11}
L.~F. {Roberts}, D.~{Kasen}, W.~H. {Lee}, E.~{Ramirez-Ruiz}, {\it \apjBib\/}
  {\bf 736}, L21 (2011).

\bibitem{Barnes16}
J.~{Barnes}, D.~{Kasen}, M.-R. {Wu}, G.~{Mart{\'{\i}}nez-Pinedo}, {\it
  \apjBib\/} {\bf 829}, 110 (2016).

\bibitem{Lee07}
W.~H. {Lee}, E.~{Ramirez-Ruiz}, {\it New Journal of Physics\/} {\bf 9}, 17
  (2007).

\bibitem{Berger14}
E.~{Berger}, {\it \araaBib\/} {\bf 52}, 43 (2014).

\bibitem{Freiburghaus99}
C.~{Freiburghaus}, S.~{Rosswog}, F.-K. {Thielemann}, {\it \apjBib\/} {\bf 525},
  L121 (1999).

\bibitem{Lattimer00}
J.~M. {Lattimer}, M.~{Prakash}, {\it \physrepBib\/} {\bf 333}, 121 (2000).

\bibitem{Dalal06}
N.~{Dalal}, D.~E. {Holz}, S.~A. {Hughes}, B.~{Jain}, {\it \prdBib\/} {\bf 74},
  063006 (2006).

\bibitem{GCN21509}
{LIGO/Virgo collaboration}, {\it GRB Coordinates Network\/} {\bf 21509} (2017).

\bibitem{GCN21513}
{LIGO/Virgo collaboration}, {\it GRB Coordinates Network\/} {\bf 21513} (2017).

\bibitem{GCN21506}
{GBM-LIGO}, {\it GRB Coordinates Network\/} {\bf 21506} (2017).

\bibitem{GCN21507}
{INTEGRAL}, {\it GRB Coordinates Network\/} {\bf 21507} (2017).

\bibitem{Gehrels16}
N.~{Gehrels}, {\it et~al.\/}, {\it \apjBib\/} {\bf 820}, 136 (2016).

\bibitem{SI}
Materials and methods are available as supplementary materials.

\bibitem{Jones09}
D.~H. {Jones}, {\it et~al.\/}, {\it \mnrasBib\/} {\bf 399}, 683 (2009).

\bibitem{Freedman01}
W.~L. {Freedman}, {\it et~al.\/}, {\it \apjBib\/} {\bf 553}, 47 (2001).

\bibitem{Drout17}
{Drout et~al.}, {\it Science, this issue 10.1126/science.aaq0049\/}  (2017).

\bibitem{Pan17}
{Pan et~al.}, {\it submitted to ApJL\/}  (2017).

\bibitem{Shappee17}
{Shappee et~al.}, {\it Science, this issue 10.1126/science.aaq0186\/}  (2017).

\bibitem{GCN21529}
{One-Meter Two-Hemisphere (1M2H) collaboration}, {\it GRB Coordinates
  Network\/} {\bf 21529} (2017).

\bibitem{GCN21551}
{Simon et~al.}, {\it GRB Coordinates Network\/} {\bf 21551} (2017).

\bibitem{MMA}
{Abbott et~al.}, {\it submitted to ApJL\/}  (2017).

\bibitem{GCN21557}
{Foley}, {\it GRB Coordinates Network\/} {\bf 21557} (2017).

\bibitem{Siebert17}
{Siebert et~al.}, {\it accepted by ApJL\/}  (2017).

\bibitem{Smartt16}
S.~J. {Smartt}, {\it et~al.\/}, {\it \mnrasBib\/} {\bf 462}, 4094 (2016).

\bibitem{Soares-Santos16}
M.~{Soares-Santos}, {\it et~al.\/}, {\it \apjBib\/} {\bf 823}, L33 (2016).

\bibitem{Schlafly11}
E.~F. {Schlafly}, D.~P. {Finkbeiner}, {\it \apj\/} {\bf 737}, 103 (2011).

\bibitem{GCN21536}
{Foley et~al.}, {\it GRB Coordinates Network\/} {\bf 21536} (2017).

\bibitem{White11}
D.~J. {White}, E.~J. {Daw}, V.~S. {Dhillon}, {\it Classical and Quantum
  Gravity\/} {\bf 28}, 085016 (2011).

\bibitem{Phinney91}
E.~S. {Phinney}, {\it \apjBib\/} {\bf 380}, L17 (1991).

\bibitem{Belczynski02}
K.~{Belczynski}, V.~{Kalogera}, T.~{Bulik}, {\it \apjBib\/} {\bf 572}, 407
  (2002).

\bibitem{Singer16}
L.~P. {Singer}, {\it et~al.\/}, {\it \apjBib\/} {\bf 829}, L15 (2016).

\bibitem{LDSS}
http://www.lco.cl/telescopes-information/magellan/instruments/ldss-3

\bibitem{Persson13}
S.~E. {Persson}, {\it et~al.\/}, {\it \paspBib\/} {\bf 125}, 654 (2013).

\bibitem{NOAO}
NOAO Archive, http://archive.noao.edu/

\bibitem{Diehl08}
H.~T. {Diehl}, {\it et~al.\/}, {\it High Energy, Optical, and Infrared
  Detectors for Astronomy III\/} (2008), vol. 7021 of {\it \procspie\/}, p.
  702107.

\bibitem{Joye03}
W.~A. {Joye}, E.~{Mandel}, {\it Astronomical Data Analysis Software and Systems
  XII\/}, H.~E. {Payne}, R.~I. {Jedrzejewski}, R.~N. {Hook}, eds. (2003), vol.
  295 of {\it Astronomical Society of the Pacific Conference Series\/}, p. 489.

\bibitem{MPC}
Minor Planet Center, http://minorplanetcenter.net

\bibitem{TNS}
Transient Name Server, https://wis-tns.weizmann.ac.il

\bibitem{Rest05:photpipe}
A.~{Rest}, {\it et~al.\/}, {\it \apjBib\/} {\bf 634}, 1103 (2005).

\bibitem{Rest14}
A.~{Rest}, {\it et~al.\/}, {\it \apjBib\/} {\bf 795}, 44 (2014).

\bibitem{Bertin02}
E.~{Bertin}, {\it et~al.\/}, {\it Astronomical Data Analysis Software and
  Systems XI\/}, D.~A. {Bohlender}, D.~{Durand}, T.~H. {Handley}, eds. (2002),
  vol. 281 of {\it Astronomical Society of the Pacific Conference Series\/}, p.
  228.

\bibitem{Skrutskie06}
M.~F. {Skrutskie}, {\it et~al.\/}, {\it \ajBib\/} {\bf 131}, 1163 (2006).

\bibitem{Alard00}
C.~{Alard}, {\it \aapsBib\/} {\bf 144}, 363 (2000).

\bibitem{Becker15}
A.~{Becker}, {HOTPANTS: High Order Transform of PSF ANd Template Subtraction},
  Astrophysics Source Code Library, 1504.004 (2015).

\bibitem{Schechter93}
P.~L. {Schechter}, M.~{Mateo}, A.~{Saha}, {\it \paspBib\/} {\bf 105}, 1342
  (1993).

\bibitem{Chambers16}
K.~C. {Chambers}, {\it et~al.\/}, {\it ArXiv}, 1612.05560 (2016).

\bibitem{Flewelling16}
H.~A. {Flewelling}, {\it et~al.\/}, {\it ArXiv}, 1612.05243 (2016).

\bibitem{Magnier16}
E.~A. {Magnier}, {\it et~al.\/}, {\it ArXiv}  1612.05242 (2016).

\bibitem{Waters16}
C.~Z. Waters,  {\it et~al.\/}, {\it ArXiv}, 1612.05245 (2016).

\bibitem{Scolnic15}
D.~{Scolnic}, {\it et~al.\/}, {\it \apjBib\/} {\bf 815}, 117 (2015).

\bibitem{Kilpatrick17}
{Kilpatrick et~al.}, {\it Science, this issue 10.1126/science.aaq0073\/}
  (2017).

\bibitem{Murguia-Berthier17}
{Murguia-Berthier et~al.}, {\it submitted to ApJL\/}  (2017).

\bibitem{Foley12:09ig}
R.~J. {Foley}, {\it et~al.\/}, {\it \apjBib\/} {\bf 744}, 38 (2012).

\bibitem{Pan15:13dy}
Y.-C. {Pan}, {\it et~al.\/}, {\it \mnrasBib\/} {\bf 452}, 4307 (2015).

\bibitem{Shappee16}
B.~J. {Shappee}, {\it et~al.\/}, {\it \apjBib\/} {\bf 826}, 144 (2016).

\bibitem{Li11:rate2}
W.~{Li}, {\it et~al.\/}, {\it \mnrasBib\/} {\bf 412}, 1441 (2011).

\bibitem{Mould90}
J.~{Mould}, {\it et~al.\/}, {\it \apjBib\/} {\bf 353}, L35 (1990).

\end{thebibliography}
\end{document}